\documentclass[aps,prfluids,groupedaddress]{revtex4-1}

\usepackage{graphicx}
\usepackage{amssymb,amsmath}
\usepackage{natbib}
\usepackage{bm}

\newcommand{\Ra}{\mathrm{Ra}}
\newcommand{\Pra}{\mathrm{Pr}}
\newcommand{\Ek}{\mathrm{Ek}}
\newcommand{\Nu}{\mathrm{Nu}}

\begin{document}

\title{Turbulent rotating convection confined in a slender cylinder: the sidewall circulation}
\author{Xander M. de Wit}
\author{Andr\'es J. Aguirre Guzm\'an}
\author{Matteo Madonia}
\author{Jonathan S. Cheng}
\altaffiliation[Current address: ]{Department of Mechanical Engineering, University of Rochester, Rochester, NY 14623, USA}
\author{Herman J. H. Clercx}
\author{Rudie P. J. Kunnen}
\email{r.p.j.kunnen@tue.nl}
\affiliation{Fluids \& Flows group, Department of Applied Physics and J. M. Burgers Center for Fluid Dynamics, Eindhoven University of Technology, P.O. Box 513, 5600 MB Eindhoven, The Netherlands}

\date{\today}

\begin{abstract}
Recent studies of rotating Rayleigh--B\'enard convection at high rotation rates and strong thermal forcing have shown a significant discrepancy in total heat transport between experiments on a confined cylindrical domain on the one hand and simulations on a laterally unconfined periodic domain on the other. This paper addresses this discrepancy using direct numerical simulations on a cylindrical domain. An analysis of the flow field reveals a region of enhanced convection near the wall, the sidewall circulation. The sidewall circulation rotates slowly within the cylinder in anticyclonic direction. It has a convoluted structure, illustrated by mean flow fields in horizontal cross-sections of the flow where instantaneous snapshots are compensated for the orientation of the sidewall circulation before averaging. Through separate analysis of the sidewall region and the inner bulk flow, we find that for higher values of the thermal forcing the heat transport in the inner part of the cylindrical domain, outside the sidewall circulation region, coincides with the heat transport on the unconfined periodic domain. Thus the sidewall circulation accounts for the differences in heat transfer between the two considered domains, while in the bulk the turbulent heat flux is the same as that of a laterally unbounded periodic domain. Therefore, experiments, with their inherent confinement, can still provide turbulence akin to the unbounded domains of simulations, and at more extreme values of the governing parameters for thermal forcing and rotation. We also provide experimental evidence for the existence of the sidewall circulation that is in close agreement with the simulation results.
\end{abstract}

\maketitle

\section{Introduction\label{ch:intro}}
Rotating Rayleigh--B\'enard convection (RRBC) --- the flow in a horizontal layer of fluid confined between two rotating plates, heated from below and cooled from above --- is a conceptually simple flow problem that captures the essence of the interplay between buoyant forcing and planetary rotation that occurs in many geophysical and astrophysical flows~\cite{cajk18}, for example, atmospheric and oceanic convection, Earth's liquid-metal outer core, the interior of the giant gas planets and the convective outer layer of the Sun. The input parameters describing this flow problem are the Rayleigh ($\Ra$), Prandtl ($\Pra$) and Ekman ($\Ek$) numbers, defined as
\begin{equation}
\Ra=\frac{g\alpha\Delta T\, H^3}{\nu\kappa} \, , \quad \Pra=\frac{\nu}{\kappa} \, , \quad \Ek=\frac{\nu}{2\Omega H^2}\, ,
\end{equation}
where~$g$ denotes gravitational acceleration, $\alpha$, $\nu$ and~$\kappa$ are the thermal expansion coefficient, kinematic viscosity and thermal diffusivity of the fluid, respectively, $\Delta T$ is the temperature difference applied between the bottom and top plates which are placed a distance~$H$ apart, and~$\Omega$ is the angular frequency of rotation. Here we shall only consider systems in which gravity is parallel to the axis of rotation, while the bottom and top plates are perpendicular to the rotation axis. The Rayleigh number describes the magnitude of the thermal forcing, the Prandtl number characterizes the working fluid while the Ekman number expresses the ratio of viscous to Coriolis forces.

A popular geometry for the study of RRBC using numerical simulation is the horizontally periodic plane layer, for both direct numerical simulations (DNSs)~\cite{jlmw96,kcg06,ksa12,ghj14,fsp14,sljvcrka14,kopvl16,fgk19} and simulations of asymptotically reduced models~\cite{sjkw06,jkrv12,jrgk12,sljvcrka14,pjms16}. The main advantage is that no lateral constraint needs to be applied: the use of sidewalls could affect the flow development by choice of shape (e.g. square cuboid or cylinder) as well as by choice of boundary condition (no-slip or stress-free; zero-heat-flux or constant temperature). In contrast, experiments must resort to lateral confinement, where an upright cylinder is by far the most popular geometry~\cite{r69,ve02,ksnha09,kgc10jfm,csrgka15,rakc18}. In that case a parameter describing the geometry is required, for example the diameter-to-height aspect ratio~$\Gamma=D/H$. In fact, to push the performance of experiments towards planetary conditions the height of convection experiments has grown, while the diameter remains rather small~\cite{csrgka15,cajk18,cmak19}.

There are several issues specific to experiments that can occur in such systems. Examples include centrifugal buoyancy~\cite{ha18,ha19} and so-called non-Oberbeck--Boussinesq effects~\cite{hs14pof}. But it turns out that the choice of domain can also have profound effects.

\begin{figure}
\includegraphics[width=0.5\textwidth]{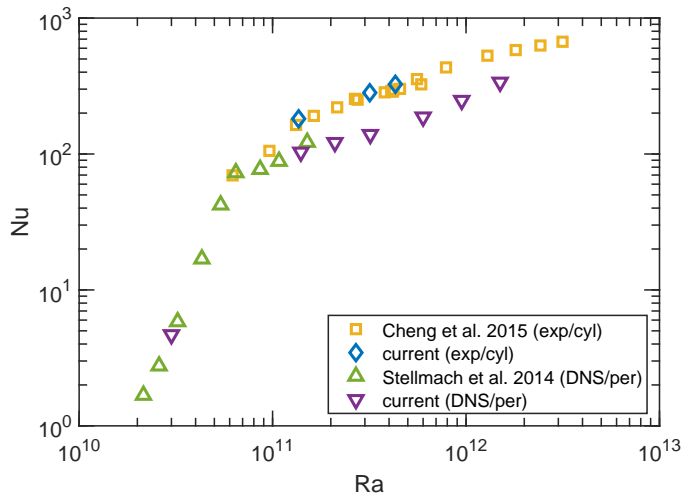}
\caption{\label{fi:nusselt_discr}Comparison of Nusselt number results from experiments in a cylinder (Cheng et al. 2015~\cite{csrgka15} and the current work) and DNS on a horizontally periodic domain (Stellmach et al. 2014~\cite{sljvcrka14} and the current work) as a function of the Rayleigh number. All simulations adopt constant~$\Ek=10^{-7}$ and~$\Pra\approx 5$ for water.}
\end{figure}

The starting point for the current investigation is a comparison of results from direct numerical simulations in a horizontally periodic domain and from experiments in an upright cylinder. The efficiency of the convective heat transfer is commonly expressed as the Nusselt number~$\Nu=q/q_\mathrm{cond}$, the total heat flux~$q$ normalized by the conductive flux~$q_\mathrm{cond}=k\Delta T/H$; $k$ is the thermal conductivity of the fluid. In Figure~\ref{fi:nusselt_discr} we compare the Nusselt number from experiments and direct numerical simulations in horizontally periodic domains. All datapoints are for water ($\Pra\approx 5$) and at constant~$\Ek=10^{-7}$. The two experiments included in the plot, TROCONVEX (Eindhoven)~\cite{cajk18,cmak19} and RoMag (UCLA)~\cite{csrgka15}, align nicely, as do the two numerical studies (Stellmach et al.~\cite{sljvcrka14} and the current study). However, for~$\Ra>10^{11}$ there is a gap in~$\Nu$ between numerical and experimental data (i.e. cylinder and periodic layer), roughly a factor two in magnitude. This discrepancy is unexpected: as the cylindrical domains in experiments were always wide enough to accommodate many times the characteristic horizontal convective scale~\cite{cajk18}, no confinement effects were anticipated.

In this paper we want to explain this discrepancy using a set of DNSs at the same operating conditions as the experiments and now in the same domain shape and boundary conditions, i.e. a cylinder of aspect ratio~$\Gamma=1/5$. We find a good agreement between DNS and experiments in the same domain. The discrepancy in heat transfer between the two domains used for DNS (i.e. cylinder and horizontally periodic layer) is shown to be caused by a strong circulation that is formed on the sidewall and significantly contributes to the overall heat flux. However, when excluding the sidewall region, we recover identical heat transfer for the bulk of the cylinder and the horizontally periodic layer.

A recent paper~\cite{zghwzaewbs19} describes a similar flow feature. While further in-depth comparison is required given that the parameter values used are quite far apart, it is expected that the same feature of confined rotating convection is studied. Hence we will identify similarities and differences between our findings and those reported in~\cite{zghwzaewbs19}. The two studies give complementary views on this interesting flow structure with obvious consequences for the interpretation of experiments.

\section{\label{ch:methods}Numerical and experimental methods}
The governing equations of turbulent rotating convection in the Boussinesq approximation are~\cite{c61}
\begin{subequations}
\begin{equation}
\frac{\partial\bm{u}'}{\partial t'} + (\bm{u}'\bm{\cdot\nabla}') \bm{u}' + \left(\frac{\Pra}{\Ra}\right)^{1/2}\frac{1}{\Ek}\,\hat{\bm{z}} \bm{\times u}'= -\bm{\nabla}' p' + \left(\frac{\Pra}{\Ra}\right)^{1/2} \nabla'^2\bm{u}' + T' \hat{\bm{z}} \, ,
\label{eq:ns}
\end{equation}
\begin{equation}
\frac{\partial T'}{\partial t'} + (\bm{u}' \bm{\cdot\nabla}') T' = \frac{1}{(\Ra\Pra)^{1/2}} \nabla'^2 T' \, ,
\label{eq:heat}
\end{equation}
\begin{equation}
\bm{\nabla}' \bm{\cdot u}' = 0\, ,
\label{eq:incomp}
\end{equation}
\end{subequations}
describing the evolution of the velocity~$\bm{u}$ and temperature~$T$ in time~$t$, where~$p$ is the pressure and~$\hat{\bm{z}}$ is the vertical unit vector. Primed symbols, e.g. $\bm{u}'$, denote nondimensionalized quantities. We  use the cell height~$H$, temperature difference~$\Delta T$ and convective velocity~$U=(g\alpha\Delta T H)^{1/2}$ for nondimensionalization.

The numerical simulations presented here are done using two codes, both using the principal setup of the Verzicco code~\cite{vo96}, well-known in the convection community.

The cylinder code is an updated version of the original Verzicco code~\cite{vo96} with extensions for better parallel performance. The code discretizes the governing equations in cylindrical coordinates with second-order accurate finite-difference approximations as further detailed in~\cite{vo96,vc03}. The domain is a slender upright cylinder with~$\Gamma=1/5$. All walls are no-slip. Constant-temperature conditions are applied at the top ($T/\Delta T=0$) and bottom ($T/\Delta T=1$) plates, while the sidewall is adiabatic. Five different simulations have been performed, at~$\Ra\in\left\{0.5,\ 0.7,\ 1.4,\ 3.2,\ 4.3\right\}\times 10^{11}$ with constant~$\Ek=10^{-7}$ and~$\Pra=5.2$. The same grid has been used in all cases, using~$769\times 351\times 1025$ grid points in the azimuthal, radial and axial directions, respectively. The points are evenly spaced in the azimuthal direction. In radial and axial directions the grids become finer near the walls, more so in the axial direction given that the Ekman boundary layers forming near bottom and top plates are significantly thinner than the sidewall boundary layer. We find \emph{a posteriori} that there are~$15$ grid points within the Ekman layers, which is enough to accurately resolve them. Bulk resolution should be compared to the Batchelor scale~$\eta_T$, representing the smallest lengthscale that should be resolved~\cite{vc03,kgc10jfm}. We see that the largest gridpoint separation never exceeds four times the local~$\eta_T$, as deemed adequate in~\cite{vc03}, but for the largest-$\Ra$ simulation where the maximal gridpoint separation always remains below five times the local~$\eta_T$. Adequacy of the grid is further confirmed by convergence to within a few percent of five ways of measuring~$\Nu$ (the average temperature gradient on the bottom and top plates, direct integration of the convective flux over the volume, and formulations based on the exact relations for the dissipation rates of kinetic energy and thermal variance, respectively~\cite{ss90}) and the excellent agreement with the experimental results (Figure~\ref{fi:nusselt_full}).

The code for the horizontally periodic domain is a multigrid adaptation of the second-order accurate finite-difference discretization of the governing equations, now on a Cartesian grid, further detailed in~\cite{oyplv15}. The refinement procedure entails that the primary grid is refined by a refinement factor~$n=2$; the temperature field is evaluated on the refined mesh, while for the velocity field the coarser primary grid is sufficient. The bottom and top plates are no-slip and isothermal, as before. The horizontal periodicity length is set to~$10L_c$, where~$L_c=4.82\Ek^{1/3}$ is the most unstable wavelength for convective onset in rapidly rotating convection~\cite{c61}. This procedure is now commonly adopted, e.g.~\cite{sjkw06,sljvcrka14,kopvl16}. In all cases there are at least~$12$ gridpoints for velocity found within the Ekman boundary layers (meaning at least twice that number of gridpoints for temperature). For bulk resolution the multigrid approach helps a lot: \emph{a posteriori} we find that~$\Delta z_u/\eta <3.5$ and~$\Delta z_T/\eta_T<4$ in all cases, where~$\Delta z_u$ is the vertical grid spacing for the velocity field and~$\Delta z_T$ for the temperature field, and~$\eta$ is the Kolmogorov length (the smallest flow lengthscale). These validations combined with the checks using the various definitions of computing~$\Nu$ reassure us of the adequacy of the grid to fully resolve this flow.

The experimental results are from TROCONVEX, the setup in Eindhoven. It is described in~\cite{cajk18,cmak19}. It is a slender cylinder of diameter~$D=0.39\;\mathrm{m}$ consisting of four segments, allowing for different working heights. The measurements in this work are taken in a cylinder of~$H=2\;\mathrm{m}$ with~$\Gamma=0.195$. The cylinder is enclosed by copper plates. The bottom plate is electrically heated while the top plate is kept at a constant temperature by circulation of cooling water from a chiller/thermostatic bath combination. The cylinder is encapsulated in insulation foam surrounded by active heat shields that take on the same temperature measured by thermistors in the sidewall; the entire height is divided into five different shields, each connected to two thermistors measuring the temperature at that height. Similarly, there is an active heater below the bottom plate that adapts to the same temperature as the bottom plate. This arrangement minimizes outward conductive losses and is the standard for precise heat-flux measurements. Note that the sidewall thermistors turn out to be interesting for flow diagnostics too. For the three measurements considered here~$\Ek=10^{-7}$ and the mean temperature is~$T_m=31.0^\circ\mathrm{C}$, so~$\Pra=5.2$. To ease comparison we shall also nondimensionalize the experimental data using the cell height~$H=2\;\mathrm{m}$, applied temperature difference~$\Delta T=\{0.63,\ 1.47,\ 1.99\}^\circ\mathrm{C}$ and convective time~$\tau_c=H/U=\{32.1,\ 21.0,\ 18.1\}\;\mathrm{s}$ for~$\Ra=\{1.4,\ 3.2,\ 4.3\}\times 10^{11}$, respectively. From here on all quantities are expressed in these units. Note that we renormalize the radial coordinate with the radius~$R=D/2$, so that~$r/R=1$ on the sidewall, for ease of interpretation.

\section{\label{ch:results}Results}
\subsection{\label{ch:instantaneous}Instantaneous snapshots}
\begin{figure} 
\includegraphics[width=0.5\textwidth]{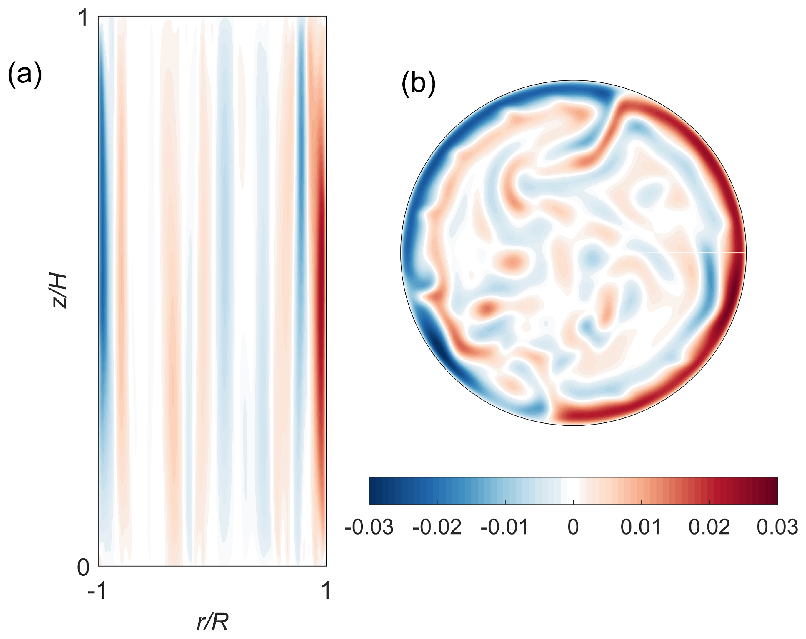}\includegraphics[width=0.5\textwidth]{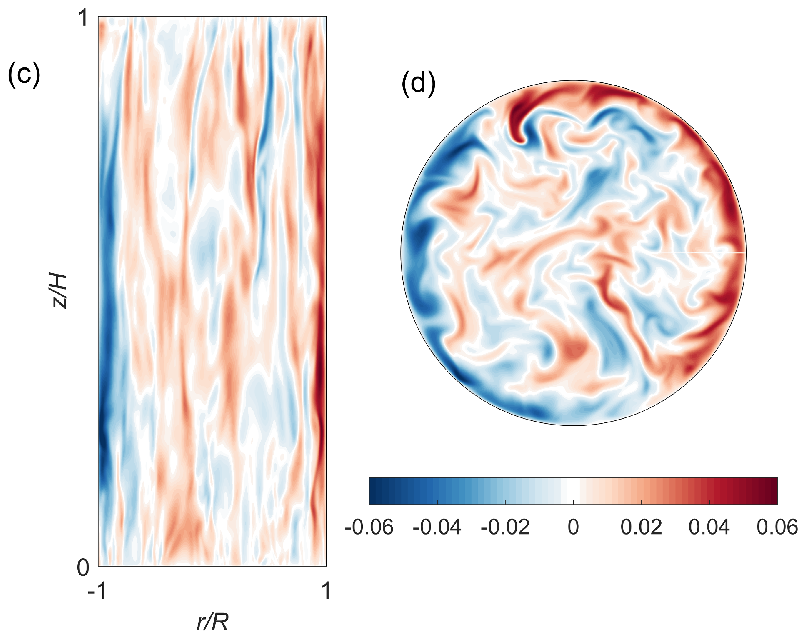}
\caption{\label{fi:inst_cross_secs}Snapshots of vertical velocity~$u_z/U$ from the simulations at (a,b) $\Ra=5\times 10^{10}$ and (c,d) $\Ra=4.3\times 10^{11}$. Panels (a,c) display vertical cross-sections where the horizontal direction has been stretched by a factor two for clarity; panels (b,d) are horizontal cross-sections at~$z/H=0.5$.}
\end{figure}

From the cylinder DNS we can get detailed information on the flow field. To start the exploration, we show instantaneous cross-sections displaying the vertical component of velocity for~$\Ra=5.0\times 10^{10}$ and~$4.3\times 10^{11}$ in Figure~\ref{fi:inst_cross_secs}. Horizontal cross-sections at~$z/H=0.5$ and vertical cross-sections are included. The Rayleigh number has a significant effect on the overall flow field: at the lower~$\Ra$ rotation is stronger than buoyancy and vertical alignment along the rotation axis is enforced, while at the higher~$\Ra$ there is more three-dimensional dynamics. This effect is well-known from previous studies~\cite{jrgk12,sljvcrka14,csrgka15} where the overall flow phenomenology was characterized. At this~$\Ek$, for~$\Ra\lesssim 1.1\times 10^{11}$ (based on the empirical relation Eq.~(19) of~\cite{csrgka15}) we expect to recover the convective-Taylor-column (CTC) regime, with plumes taking the shape of columnar vortical chimneys connecting both boundary layers and shielded with patches of opposite vorticity. The `torsional' nature of the columns~\cite{pkhm08,gjwk10}, with vertical vorticity changing sign from cyclonic to anticyclonic while crossing the vertical extent of the domain, is confirmed with plots of vertical vorticity (not shown here). The simulation at~$\Ra=7.0\times 10^{10}$ also renders a CTC-type flow. At higher~$\Ra\gtrsim 1.5\times 10^{11}$ the vertical coherence is relaxed, but still the dominant force balance is the geostrophic balance between Coriolis and pressure gradient. This flow regime is referred to as plumes~\cite{jrgk12,sljvcrka14,csrgka15}. At~$\Ra=1.4\times 10^{11}$ (not shown) some columns can be found but the columnar structure is falling apart, hence a transitional state. For the two highest-$\Ra$ cases considered here we recover a flow as displayed in Figure~\ref{fi:inst_cross_secs}(c,d), belonging to either plumes or geostrophic-turbulence regimes~\cite{sjkw06,jrgk12,csrgka15,cajk18} that are difficult to discern by eye (and which is not our current objective).

However, all simulations share a predominant flow feature: there is a region of large vertical velocity near the sidewall. This represents a structure of fluid flowing up along one side of the wall and down along the opposite side. It can be observed throughout almost the full vertical extent of the domain; it is only close to the Ekman layers that the prominence of this flow is reduced. We will refer to this flow feature as the sidewall circulation. Zhang et al.~\cite{zghwzaewbs19} coined it the boundary zonal flow (BZF); here we shall not adopt that name given the different results from the analysis of the sidewall flow feature that we will indicate later, which could also indicate different structures. Nevertheless, here, as in~\cite{zghwzaewbs19}, the sidewall circulation contributes significantly to the overall heat transfer.

\subsection{\label{ch:size}Size of the sidewall circulation}
\begin{figure}
\includegraphics[width=\textwidth]{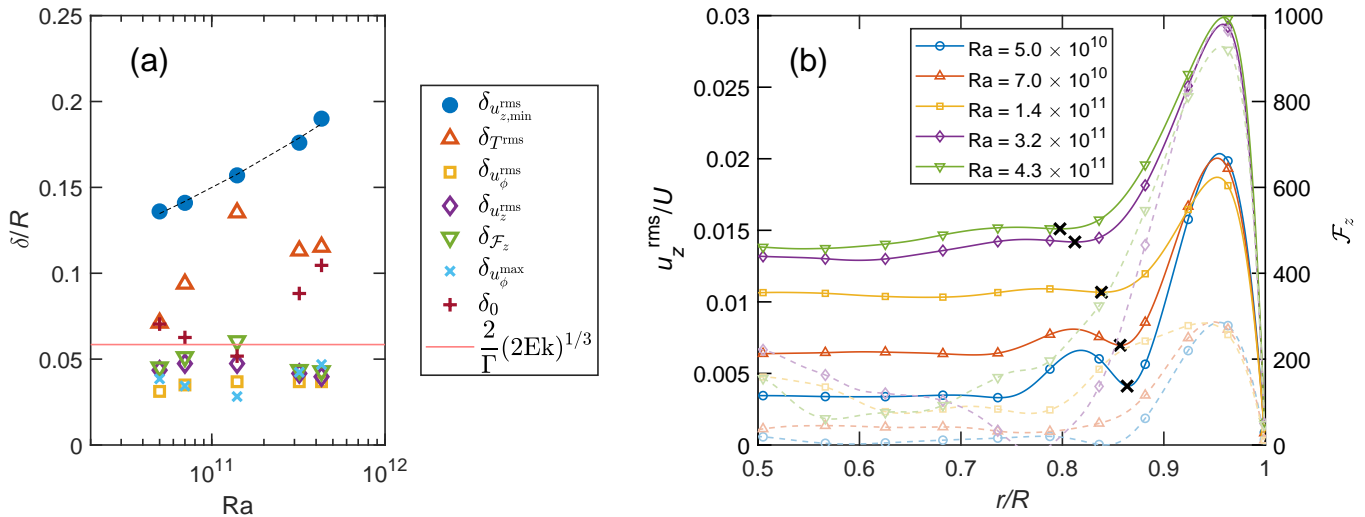}
\caption{\label{fi:blscales}(a) Various boundary-layer scales evaluated at different~$\Ra$. The definitions are introduced in the text; the horizontal line indicates the boundary layer thickness~$\delta_S/R=(2/\Gamma)(2\Ek)^{1/3}$ reported to be a good fit to the sidewall boundary layer thickness in previous studies~\cite{kgc10jfm,ksoshc11,kcc13,kch13}. The black dashed line is a power-law fit~$\delta_{u_{z,\mathrm{min}}^\mathrm{rms}}/R=(3\pm 1)\times 10^{-3}\cdot\Ra^{0.15\pm 0.02}$. (b) Radial dependence of~$u_z^\mathrm{rms}$ (solid lines with dark shading; left ordinate) and~$\mathcal{F}_z$ (dashed lines with light shading; right ordinate) at~$z/H=0.5$. The black crosses indicate the positions of the local minima in~$u_z^\mathrm{rms}$, the distance of these points to the sidewall is called~$\delta_{u_{z,\mathrm{min}}^\mathrm{rms}}$.}
\end{figure}

The definition of the size of a boundary layer like this is quite ambiguous. Different definitions may be used, see e.g. the various definitions of the thermal boundary layer thickness as compared in~\cite{jrgk12}. Zhang et al.~\cite{zghwzaewbs19} have introduced four definitions that we will also apply (all evaluated at mid-height~$z/H=0.5$):
\begin{itemize}
\item $\delta_0$, where the time-and-azimuthally averaged azimuthal velocity~$\left<u_\phi\right>_{t,\phi}$ is zero;
\item $\delta_{u_\phi^\mathrm{max}}$, where~$\left<u_\phi\right>_{t,\phi}$ reaches its maximum;
\item $\delta_{\mathcal{F}_z}$, where the normalized local vertical heat flux~$\mathcal{F}_z=[u_z(T-T_m)-\kappa\partial T/\partial z]/(\kappa\Delta T/H)$ attains its maximum, with~$T_m=(T_\mathrm{bottom}+T_\mathrm{top})/2$ the mean temperature of bottom and top plates.
\item $\delta_{u_z^\mathrm{rms}}$, where the root-mean-square (rms) vertical velocity is maximal;
\end{itemize}
Additionally, we consider three more boundary layer scales:
\begin{itemize}
\item $\delta_{u_\phi^\mathrm{rms}}$, location of the maximal azimuthal rms velocity;
\item $\delta_{T^\mathrm{rms}}$, location of the maximal rms temperature;
\item $\delta_{u_{z,\mathrm{min}}^\mathrm{rms}}$, the location of the near-wall minimum of~$u_z^\mathrm{rms}$.
\end{itemize}
The boundary layer scales are plotted as a function of~$\Ek$ in Figure~\ref{fi:blscales}(a); We plot the radial dependence of~$u_z^\mathrm{rms}$ for the considered simulations in Figure~\ref{fi:blscales}(b) to further illustrate the definition of~$\delta_{u_{z,\mathrm{min}}^\mathrm{rms}}$. We note that the measures based on azimuthal averaging of azimuthal velocity disregard some of the more complex features of the circulation, as will be shown in Section~\ref{ch:structure}. The measures based on peak rms velocity and peak heat flux are quite similar overall and in line with earlier studies on sidewall Stewartson boundary layers in rotating cylindrical convection~\cite{kgc10jfm,ksoshc11,kcc13,kch13}, with a thickness scaling as~$\delta_S/H \sim (2\Ek)^{1/3}$ (that we cannot validate here) or~$\delta_S/R\sim (2/\Gamma)(2\Ek)^{1/3}$ and a prefactor of about~$1$. Here we choose the local minimum in~$u_z^\mathrm{rms}$ for further analysis, as we found that it captures the full near-wall heat-flux peak in Fig.~\ref{fi:blscales}(b). This boundary layer thickness changes with~$\Ra$: a power-law fit results in~$\delta_{u_{z,\mathrm{min}}^\mathrm{rms}}/R=(3\pm 1)\times 10^{-3}\cdot\Ra^{0.15\pm 0.02}$.

\subsection{\label{ch:dynamics}Dynamics of the sidewall circulation}
\begin{figure}
\includegraphics[width=\textwidth]{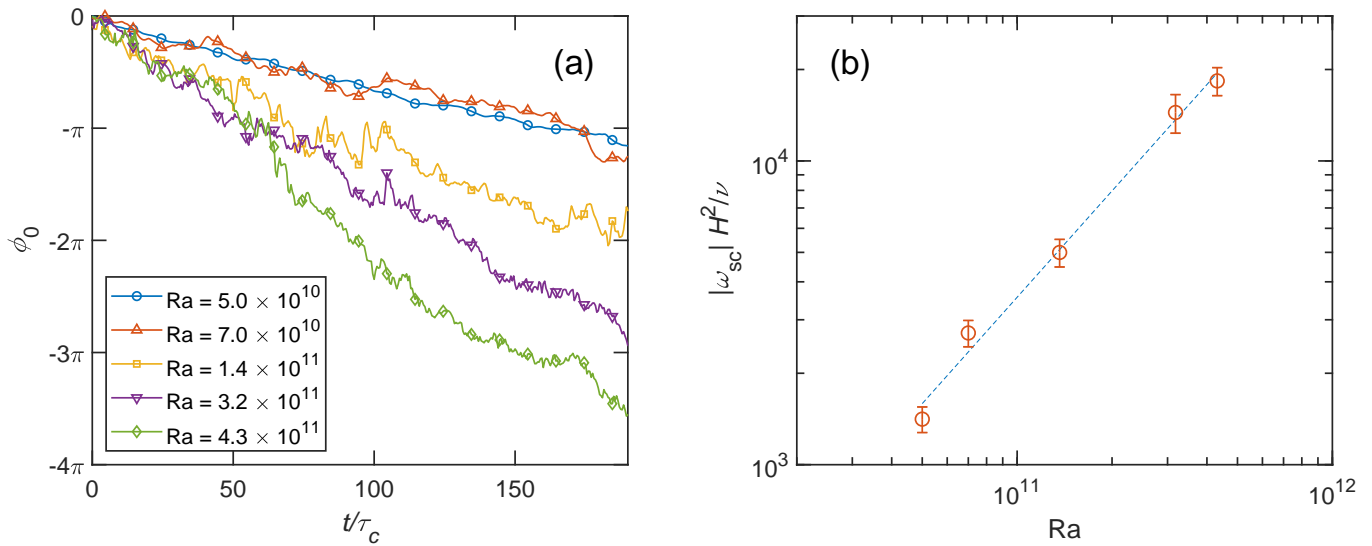}
\caption{\label{fi:offset}(a) Radial drift of the sidewall circulation: phase angle~$\phi_0$ as a function of time~$t/\tau_c$. (b) Angular velocity~$|\omega_\mathrm{sc}|$ expressed in viscous units (which are independent of~$\Ra$) as a function of~$\Ra$. The error bars represent the standard deviation of the angular velocities determined for six equal parts of the total time. The dashed line displays a power-law fit~$|\omega_\mathrm{sc}|H^2/\nu=6\times 10^{-10}\cdot\Ra^{1.16\pm 0.06}$.}
\end{figure}

From movies of the vertical velocity in a horizontal cross-sectional plane (see the Supplemental Material~\cite{suppl}) it is clear that the sidewall circulation rotates anticyclonically (i.e. counter to the rotation of the cylinder) in the co-rotating frame of reference used for the simulations. As is customary in experiments and simulations to study the large-scale circulation typical for non- and weakly rotating convection~\cite{ba06jfm,ksoshc11}, we use the azimuthal profile of vertical velocity at~$r/R=1-\tfrac{1}{2}\delta_{u_{z,\mathrm{min}}^\mathrm{rms}}/R$ and~$z/H=0.5$ from our simulations to fit a cosine function~$u_z(\phi,t)\sim\cos(\phi-\phi_0(t))$ to determine the phase angle~$\phi_0(t)$ of the sidewall circulation. In Figure~\ref{fi:offset}(a) we plot~$\phi_0$ as a function of (convective) time~$t/\tau_c$. In all cases the graphs display a trend towards negative~$\phi_0$, i.e. anticyclonic precession. To quantify the drift rate we fit lines to~$\phi_0(t)$; the slope~$\omega_\mathrm{sc}$ is negative and its dependence on~$\Ra$ is displayed in Figure~\ref{fi:offset}(b). Note that we plot~$|\omega_\mathrm{sc}|H^2/\nu$ to express the rotation rate nondimensionalized using the inverse viscous time~$1/\tau_\nu=\nu/H^2$ rather than the convective unit~$1/\tau_c=(g\alpha\Delta T/H)^{1/2}$; the viscous unit retains the same value in all simulations while~$\tau_c$ changes with~$\Ra$. The rotation rate of the sidewall circulation increases quite rapidly as~$\Ra$ is enhanced, but it must be said that in all cases this azimuthal drift is still more than two orders of magnitude slower than the rotation rate of the cylinder (which is~$1/(2\Ek)=5\times 10^6$ in the same units).

\subsection{\label{ch:structure}Orientation-compensated mean flow structure of the sidewall circulation}
\begin{figure}
\includegraphics[width=\textwidth]{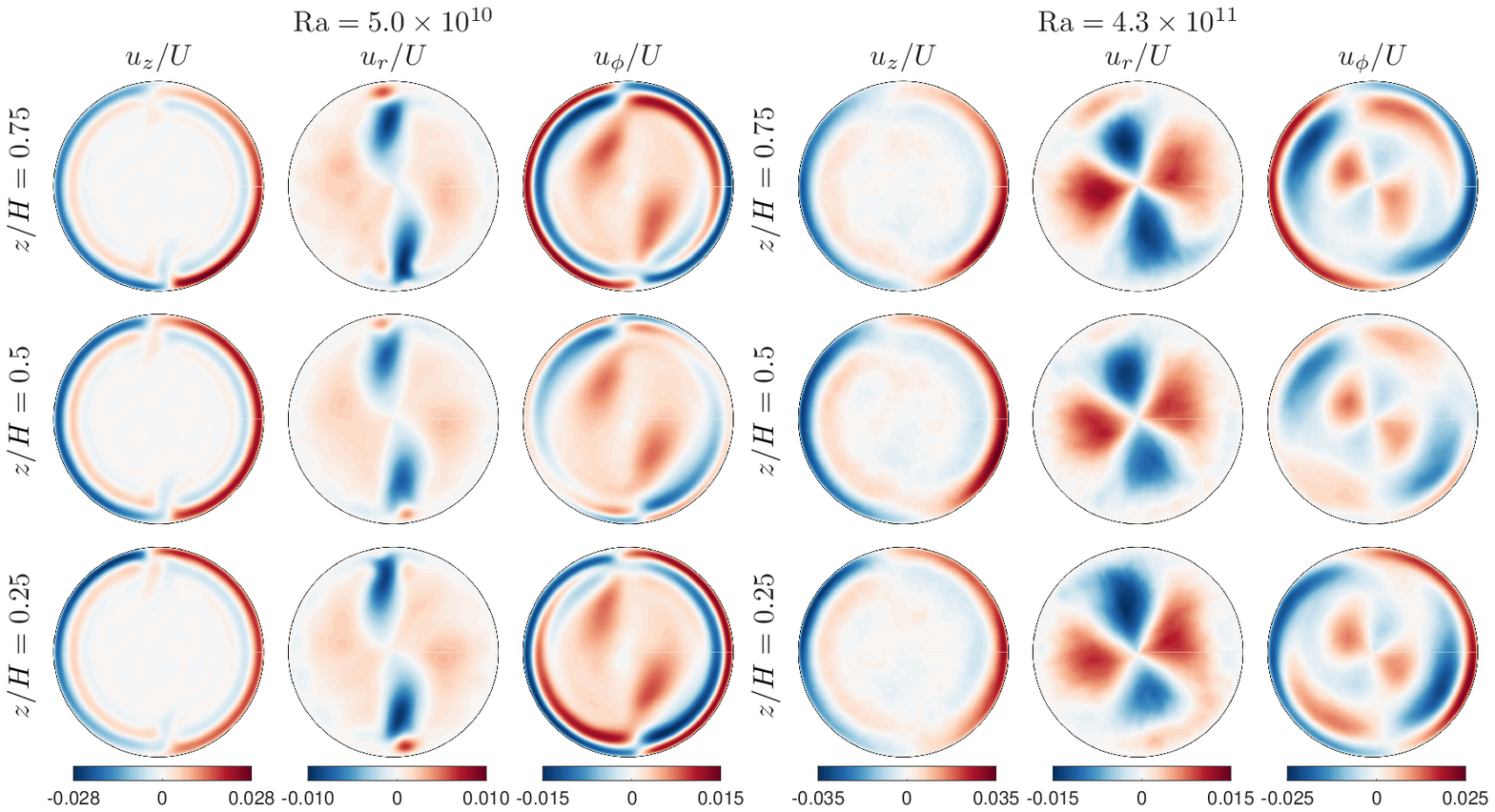}
\caption{\label{fi:phase_avg}Orientation-compensated mean velocity fields in horizontal cross-sections at heights~$z/H=0.25$, $0.5$ and~$0.75$ for two Rayleigh numbers~$5.0\times 10^{10}$ (left) and~$4.3\times 10^{11}$ (right). The vertical ($u_z$), radial ($u_r$) and azimuthal ($u_\phi$) velocity components are shown.}
\end{figure}

To understand the azimuthal drift and the enhanced near-wall heat transfer, we want to identify the circulation set up inside the sidewall circulation region. The mean flow pattern can be identified by an orientation-compensating shift followed by averaging, i.e. we rotate instantaneous horizontal cross-sections counter to their corresponding~$\phi_0(t)$ before performing an ensemble average of the flow field. The averaged snapshots provide detailed information on the mean flow pattern of the sidewall circulation, as displayed in Figure~\ref{fi:phase_avg}. The lowest and highest~$\Ra$ simulations are visualized here. Temperature is not included; temperature fluctuations around the mean for each height are distributed similarly to vertical velocity~$u_z$. What can be observed is a geostrophic bulk (i.e. velocity components are mostly independent of the vertical coordinate), surrounded by a two-layer (inner and outer), two-halves (left and right parts of the panels in Figure~\ref{fi:phase_avg}) sidewall boundary layer. Within this boundary layer the azimuthal velocity~$u_\phi$ displays up--down antisymmetry with respect to the midplane~$z/H=0.5$, while the vertical velocity~$u_z$ is vertically symmetrical. The radial velocity~$u_r$ in the boundary layer region is prominent in two spots located at the top and bottom of the panels in Fig.~\ref{fi:phase_avg}, where again approximate up--down antisymmetry can be observed. The two-halves structure is the reason why we want to avoid defining the thickness of the layer using azimuthal averaging of~$u_\phi$; there is only a minor asymmetry between the two halves, the mean of which does not correctly convey the size. Azimuthal averaging washes away all of the interesting azimuthal structure revealed here. The Supplemental Material~\cite{suppl} displays this effect for the azimuthally averaged~$u_\phi$ at~$z/H=0.5$.

In both displayed~$\Ra$ cases we observe a radial structure for~$u_z$ that consists of two layers of opposite sign akin to the Stewartson-type layers described in~\cite{kch13}. The thickness of the outermost layer, following the peak position of rms velocity, is in line with the expected thickness for Stewartson layers, as shown before. Further evidence of the layers being Stewartson layers is presented in the Supplemental Material~\cite{suppl}, where a horizontal cross-section within the bottom Ekman boundary layer reveals the action of the `corner' region near~$z/H=0$ and~$r/R=1$ of typical dimension~$\Ek^{1/2}\times\Ek^{1/2}$ (both scaled with cell height~$H$) that can be quite important for the overall flux~\cite{kch13}. The inner part of the Stewartson layer is expected to be weaker than the outer part; it is visible in Figure~\ref{fi:blscales}(b) as a smaller peak in~$u_z^\mathrm{rms}$, which is washed away by the turbulent bulk in the higher~$\Ra$ cases. In line with that, the two-layer structure in~$u_\phi$, prominent at~$\Ra=5.0\times 10^{10}$, all but disappears at~$\Ra=4.3\times 10^{11}$. Where the two halves of the layers meet there is a strong radially inward flow that can be appreciated even better from the movies in the Supplemental Material~\cite{suppl}. These jet-like eruptions set the mean flow pattern in the bulk: at the lower~$\Ra$ the jets are deflected to the right (positive~$u_\phi$) by the Coriolis acceleration; at the higher~$\Ra$ the jets collide centrally and subsequently flow outward in the perpendicular directions. Note, however, that the mean flow in the bulk is coexisting with a fluctuating turbulent field (see Sec.~\ref{ch:bulk_vs_wall_mode}).

The visualizations of Figure~\ref{fi:phase_avg} have some visual resemblance to the modes for convective onset in a cylinder~\cite{zl09}. However, at the current parameter values these onset modes are domain-filling (and not wall-localized) with azimuthal wavenumber~$m=1$ and leave out the prominent jets, leading us to conclude that what we observe are not onset modes. We currently cannot explain the structure of the sidewall circulation: its `torsional' structure of the mean~$u_\phi$ field (up--down antisymmetry) and its division into two halves. We are currently studying this in more detail with boundary-layer theory.

\subsection{\label{ch:bulk_vs_wall_mode}Contributions of bulk and sidewall circulation regions to the heat transfer}
Given that the sidewall circulation contributes significantly to the overall heat transfer, it is of interest to compare the `strength' of the sidewall circulation to the turbulence intensity. To this end we take the absolute value of~$u_z$ in the orientation-compensated mean field at~$z/H=0.5$ (Figure~\ref{fi:phase_avg}) and  average it over the sidewall-circulation region~$1-\delta_{u_{z,\mathrm{min}}^\mathrm{rms}}/R<r/R<1$ as a measure of the strength of the sidewall circulation. The bulk turbulence intensity is measured by~$u_z^\mathrm{rms}$ averaged over~$0<r/R<1-\delta_{u_{z,\mathrm{min}}^\mathrm{rms}}/R$ at the corresponding height. The resulting sidewall-circulation-to-bulk-turbulence ratio denoted by~$\Phi$ is~$\Phi\approx 3$ at the lowest~$\Ra=5.0\times 10^{10}$;~$\Phi\approx 1.5$ at~$\Ra=7.0\times 10^{10}$; it attains its minimal value~$\Phi\approx 0.5$ at~$\Ra=1.4\times 10^{11}$; then at the two highest~$\Ra$ increases again to~$\Phi\approx 0.9$. An interpretation of this ratio is that its value indicates which flow feature (turbulence or sidewall circulation) is expected to be most prominent in an instantaneous snapshot of the flow: when~$\Phi\gg 1$ it will mostly display the sidewall-circulation signature, when~$\Phi\ll 1$ we expect a foremost turbulent field. From these numbers we conclude that indeed the sidewall circulation remains significant throughout the~$\Ra$ range considered here; for the lowest~$\Ra$ it is probably the dominant dynamical feature, stronger than the bulk turbulence, while at~$\Ra=1.4\times 10^{11}$ the bulk turbulence is stronger than the sidewall circulation. This can also be appreciated from the movies in the Supplemental Material~\cite{suppl}.

\begin{figure}
\includegraphics[width=0.5\textwidth]{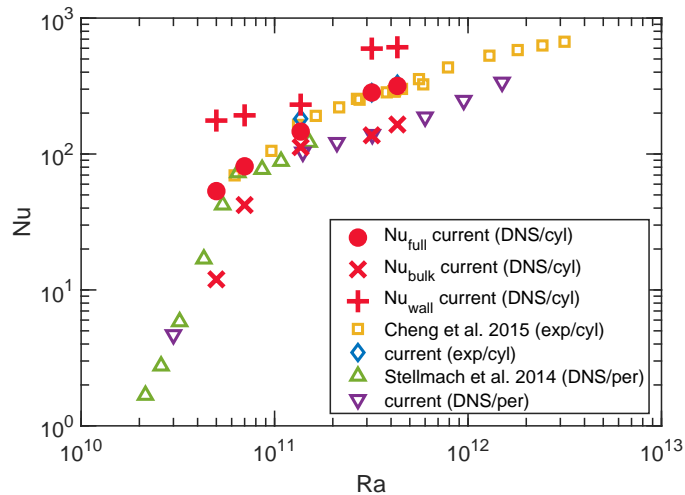}
\caption{\label{fi:nusselt_full}Nusselt number results from the current simulations. Included are~$\Nu_\mathrm{full}$ for heat flux averaged over the entire cylinder cross-section, as well as~$\Nu_\mathrm{bulk}$ and~$\Nu_\mathrm{wall}$ representing the heat flux averaged over bulk and near-wall regions, respectively. The other symbols are repeated from Figure~\ref{fi:nusselt_discr}.}
\end{figure}

The distinction between bulk and sidewall-circulation regions is also easily made in the heat transfer. We have computed the bulk (near-wall) Nusselt number~$\Nu_\mathrm{bulk}$ ($\Nu_\mathrm{wall}$) by averaging~$\mathcal{F}_z$ over~$0<r/R<1-\delta_{u_{z,\mathrm{min}}^\mathrm{rms}}/R$ ($1-\delta_{u_{z,\mathrm{min}}^\mathrm{rms}}/R<r/R<1$) at~$z/H=0.5$. These results, together with the default full-area Nusselt number~$\Nu_\mathrm{full}$, are plotted in Figure~\ref{fi:nusselt_full}. Overall, we want to mention first that~$\Nu_\mathrm{full}$ (filled red circles) nicely coincides with the experimental results, giving us trust in the current cylindrical DNS. At the two lowest~$\Ra$ considered here, part of the CTC flow regime, it can be seen that the bulk is convecting less heat than the horizontally periodic layer. We observe in the supplemental movies~\cite{suppl} that a dominant horizontal flow induced by the jets truly dominates the bulk flow, which we expect to distort and reduce the vertical natural convection. On the other hand, when considering~$\Nu_\mathrm{bulk}$ at~$\Ra\ge 1.4\times 10^{11}$ we can see a very satisfactory quantitative agreement with the results of the DNS on the horizontally periodic domain. This is a strong result that indicates that --- for the plumes and geostrophic-turbulence flow regimes at least --- the sidewall circulation is indeed only affecting the turbulent heat flux in a confined near-wall region, while the bulk displays unaffected dynamics at least with regard to the heat transfer. Obviously,~$\Nu_\mathrm{wall}$ displays the opposite trend from~$\Nu_\mathrm{bulk}$ where it reaches values higher than~$\Nu_\mathrm{full}$ as a clear sign that the near-wall region contributes more than its share in area to the overall heat flux. If~$\Ra$ would be increased beyond~$4.3\times 10^{11}$, we expect the sidewall circulation to disappear once the transition to rotation-affected flow takes place. Then the three quantities~$\Nu_\mathrm{full}$, $\Nu_\mathrm{bulk}$ and~$\Nu_\mathrm{wall}$ should be closer together.

\subsection{\label{ch:exp_evidence}Experimental evidence for existence of the sidewall circulation}
\begin{figure}
\includegraphics[width=\textwidth]{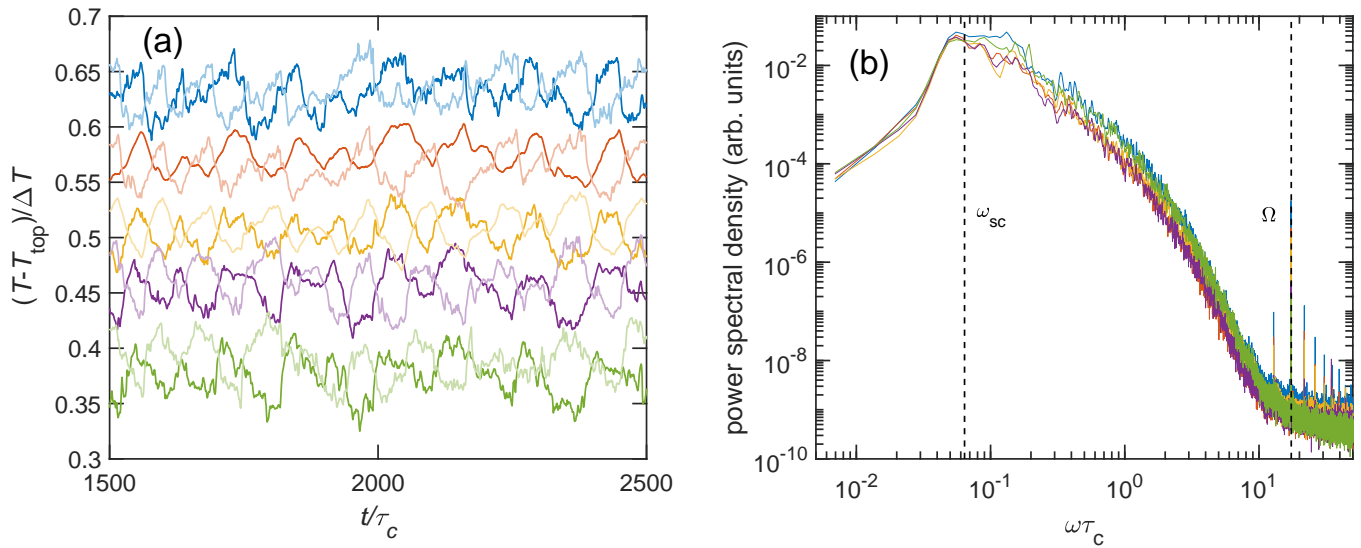}
\caption{\label{fi:walltemp_spectra}(a) Partial time traces (in convective time units) of sidewall temperature probes at (from top to bottom)~$z/H=0.1$, $0.3$, $0.5$, $0.7$ and~$0.9$. Each pair of lines with the same color but different shading displays the signals from sensors on opposite sides of the cylinder but at the same height. These measurements are taken at~$\Ra=4.3\times 10^{11}$. (b) Frequency spectra of the sidewall temperature time traces in panel (a). Only one sensor per height is included for clarity. The vertical dashed lines indicate reference angular frequencies:~$\omega_\mathrm{sc}$ from the corresponding DNS (Figure~\ref{fi:offset}) and~$\Omega$, the rotation rate of the setup.}
\end{figure}

The sidewall temperature probes provide evidence for the existence of the sidewall circulation in the experiment. In Figure~\ref{fi:walltemp_spectra}(a) we plot part of the time trace of temperature measurements of the sidewall probes; two pairs of probes on opposite sides of the cylinder at five different heights. The displayed part covers~$1000$ convective time units, i.e. about five hours out of a total segment of 24 hours measured in this experiment. Each trace displays a somewhat erratic but clearly evident oscillation around a mean value. This mean value is a function of height, a well-known result of the mean temperature gradient that develops in turbulent rotating convection~\cite{jlmw96}. There is a half-period phase difference between signals from opposite sides of the cylinder wall (darker and lighter shades of the same color), while signals on the same side but at different heights are at the same phase (darker shades align in phase, as do the lighter shades). These measurements are fully in line with the presence of a sidewall circulation as described in the earlier sections. Since we only have two thermistors per height we cannot determine the direction of the precession. But it is possible to calculate frequency spectra of the temperature time traces. These are reported in Figure~\ref{fi:walltemp_spectra}(b), where we only include one spectrum per vertical position for clarity; the others are similar. The shape of the spectrum is independent of height. The precession rate~$\omega_\mathrm{sc}$ of the sidewall circulation as inferred from the corresponding DNS is indicated with a vertical dashed line. It coincides satisfactorily with the peak of the spectra. This shows that the precession period of the sidewall circulation is the largest active time scale in this flow. We can compute similar spectra from the DNS for reference; we do not show them here given that the shorter elapsed time (only a few hundreds of~$\tau_c$) leads to inferior frequency resolution, but the peak is found at the expected position. The spectra of the two other experimental cases (shown in the Supplemental Material~\cite{suppl}) display similar results; the experiment at~$\Ra=1.4\times 10^{11}$ has a sidewall-circulation peak that is lower, in line with the observed lower intensity of the sidewall circulation in that case (Section~\ref{ch:bulk_vs_wall_mode}).

\section{\label{ch:concl}Conclusion}
In this combined numerical--experimental investigation of rapidly rotating turbulent convection in a cylindrical cell we have found a prominent sidewall circulation that provides a significant contribution to the overall heat transfer. It consists of a two-layer, two-halves structure with upward flow near the sidewall on one side and downward flow on the opposite side, while the azimuthal velocity displays a division into two radially separated bands with opposite direction of motion and a vertical torsional structure (antisymmetry with respect to the horizontal midplane). Jets consisting of intermittent bursts are ejected into the bulk from the locations where the two halves meet. This entire arrangement drifts slowly in the anticyclonic direction on timescales several orders of magnitude larger than the rotation period of the cell.

The sidewall circulation described here has similarities to the boundary zonal flow (BZF) reported by~\cite{zghwzaewbs19}, but we also find remarkable differences, e.g., the two-halves structure. We expect that most of the differences occur because of different methods of analysis. In particular the use of azimuthal averaging in~\cite{zghwzaewbs19} obscures a lot of the important properties of the sidewall circulation that we report here, due to the approximate antisymmetry of the two halves. Additionally, it is entirely possible that the differences in the values of the governing parameters are responsible for the observed differences, given that our Prandtl number is considerably larger ($\Pr=5.2$ versus~$0.8$ in~\cite{zghwzaewbs19}), our Rayleigh numbers are about two orders of magnitude larger, and our Ekman number is more than one order of magnitude smaller than the smallest~$\Ek$ considered in~\cite{zghwzaewbs19}. Finally, the cylinder aspect ratio (here~$\Gamma=1/5$, while~\cite{zghwzaewbs19} apply~$\Gamma=1/2$) is different. There is certainly need for further reconciliation of these results, by investigating the dependence of properties of the sidewall circulation or BZF on the parameter values.

This sidewall circulation contributes significantly to the overall heat transfer in a cylinder. Some care must be taken in the interpretation of these results as far as a comparison to horizontally periodic simulation domains or geophysically and astrophysically relevant geometries is intended (e.g.~\cite{csrgka15,zghwzaewbs19,cmak19}). However, our current findings hint at the sidewall circulation being a Stewartson-type boundary layer, which identifies clear avenues of reducing its relative importance: either lower~$\Ek$ to reduce its radial size, or increase the lateral dimensions of the convection cell. Nevertheless, for high enough Rayleigh number, the heat transfer properties of the bulk flow are clearly in line with those of simulated flows in a laterally periodic domain. Hence we expect the bulk flow to be unaffected by the sidewall. Experiments using stereoscopic particle image velocimetry in TROCONVEX are in preparation; with these measurements we will perform a more in-depth comparison of flow statistics between bulk and sidewall regions.

The influence of lateral confinement on rotating flow systems should never be underestimated. In this paper we have identified a sidewall circulation that exists only due to the presence of a sidewall. Furthermore, we have carried out some further (preliminary) simulations, both in a square cuboid domain (courtesy of Paolo Cifani, University of Groningen) and in a cylinder domain with stress-free sidewalls, that have also revealed the existence of a sidewall circulation analagous to the one described in this paper, but with differences in details of the structure. The combination of Ekman boundary layers and lateral confinement appears to be enough to set up a sidewall circulation. Yet, notwithstanding the prominence of the sidewall circulation, we have shown that a turbulent flow that is statistically unaffected by the sidewall circulation can still develop in the bulk of confined domains. Therefore, experiments, despite their inherent confinement, are still very valuable: they can provide bulk turbulence unaffected by sidewall effects at more extreme parameter values than can be achieved in numerical simulations, getting closer to the natural systems in geophysics and astrophysics that we want to understand and describe.

\begin{acknowledgments}
A.J.A.G., M.M., J.S.C. and R.P.J.K. have received funding from the European Research Council (ERC) under the European Union's Horizon 2020 research and innovation programme (Grant agreement No. 678634). We are grateful for the support of the Netherlands Organisation for Scientific Research (NWO) for the use of supercomputer facilities (Cartesius) under Grants No. 15462, 16467 and 2019.005. Furthermore, we thank Paolo Cifani (University of Groningen) for the preliminary simulation on the square cuboid domain.
\end{acknowledgments}

\end{document}